\documentclass[prd,nofootinbib]{revtex4} 
\usepackage{graphicx} 
\usepackage{amsmath}
\usepackage{amsfonts,amsbsy}
\usepackage{amssymb}

\def\be{\begin{equation}}
\def\ee{\end{equation}}
\def\bea{\begin{eqnarray}}
\def\eea{\end{eqnarray}}

\begin{document}

\title{\bf KNO scaling from a nearly Gaussian action for small-$x$ gluons}


\author{Adrian Dumitru$^{a,b,c}$ and Elena Petreska$^{b,c}$}
\affiliation{
$^a$ RIKEN BNL Research Center, Brookhaven National
  Laboratory, Upton, NY 11973, USA\\
$^b$ Department of Natural Sciences, Baruch College, CUNY,\\
17 Lexington Avenue, New York, NY 10010, USA\\
$^c$ The Graduate School and University Center, The City
  University of New York, 365 Fifth Avenue, New York, NY 10016, USA\\
}

\begin{abstract}
Transverse momentum integrated multiplicities in the central region of
$pp$ collisions at LHC energies satisfy Koba-Nielsen-Olesen
scaling. We attempt to relate this finding to multiplicity
distributions of soft gluons.  KNO scaling emerges if the effective
theory describing color charge fluctuations at a scale on the order of
the saturation momentum is approximately Gaussian. From an evolution
equation for quantum corrections which includes both saturation as
well as fluctuations we find that evolution with the QCD
$\beta$-function satisfies KNO scaling while fixed-coupling evolution
does not. Thus, non-linear saturation effects and running-coupling
evolution are both required in order to reproduce geometric scaling of
the DIS cross section and KNO scaling of virtual dipoles in a hadron
wave function.
\end{abstract}

\maketitle

\section{Introduction}

The color fields of hadrons boosted to the light cone are thought to
grow very strong, parametrically of order $A^+ \sim 1/g$ where
$g$ is the coupling~\cite{Mueller:1999wm}. The fields of nuclei are
enhanced further by the high density of valence charges per unit
transverse area, which is proportional to the thickness $A^{1/3}$ of a
nucleus~\cite{MV}.

In collisions of such strong color fields a large number of soft
gluons is released. Due to the genuinely non-perturbative dynamics of
the strong color fields a semi-hard ``saturation scale'' $Q_s$
emerges; it corresponds to the transverse momentum where the phase
space density of produced gluons is of order $1/\alpha_s$. The mean
multiplicity per unit rapidity in high-energy collisions is then
$\bar{n}\equiv \langle dN/dy\rangle \sim 1/\alpha_s$. Below we argue
that a semi-classical effective theory of valence color charge
fluctuations predicts that the variance of the multiplicity
distribution is of order $k^{-1}\sim {\cal O}(\alpha_s^0)$ so that the
perturbative expansion of $\bar{n}/k$ begins at order $1/\alpha_s\gg
1$. We show that in the strong field limit then a Gaussian effective
theory leads to Koba-Nielsen-Olesen (KNO)
scaling~\cite{Koba:1972ng}. This relates the emergence of KNO scaling
in $p_\perp$-integrated multiplicity distributions from high-energy
collisions to properties of soft gluons around the saturation scale.

\begin{figure}[htb]
\begin{center}
\includegraphics[width=0.5\textwidth]{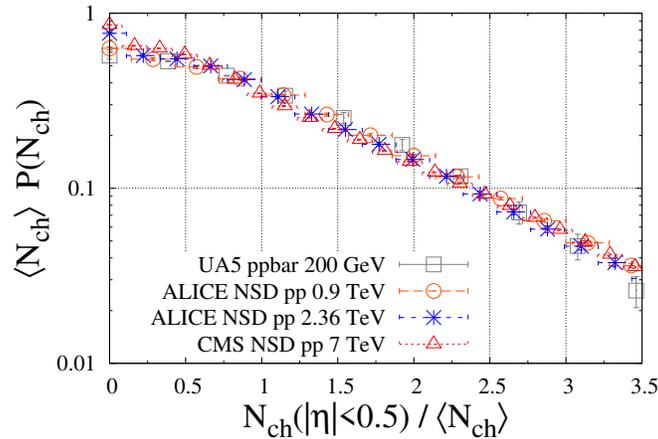}
\end{center}
\vspace*{-0.5cm}
\caption[a]{KNO scaling of charged particle multiplicity distributions
  in non-single diffractive $pp\,/\,p\overline{p}$ collisions at various
  energies as measured by the UA5~\cite{Ansorge:1988kn},
  ALICE~\cite{Aamodt:2010ft} and CMS~\cite{Khachatryan:2010nk}
  collaborations, respectively. Note that we restrict to the bulk of
  the distributions up to 3.5 times the mean multiplicity.}
\label{fig:KNO_LHCdata}
\end{figure}
The KNO scaling conjecture refers to the fact that the particle
multiplicity distribution in high-energy hadronic collisions is {\em
  universal} (i.e., energy independent) if expressed in terms of the
fractional multiplicity $z\equiv n/\bar n$. This is satisfied to a
good approximation in the central (pseudo-) rapidity region at center
of mass energies of 900~GeV and
above~\cite{Aamodt:2010ft,Khachatryan:2010nk} as shown in
fig.~\ref{fig:KNO_LHCdata}. On the other hand, UA5
data~\cite{Ansorge:1988kn} taken at $\surd s = 200$~GeV appears to
show a slightly distorted multiplicity distribution. This is
in line with the observation that at lower energies higher-order
factorial moments $G_q$ of the distribution are energy dependent and
significantly {\em different} from the reduced moments
$C_q$~\cite{ZajcPLB175}:
\be
G_q \equiv \frac{\langle n\, (n-1)\cdots (n-q+1)\rangle}{\bar{n}^q} ~~~,~~~
C_q \equiv \frac{\langle n^q\rangle}{\bar{n}^q}~~.
\ee
In fact, since the difference of $G_q$ and $C_q$ is subleading in the
density of valence charges one may interpret this finding to indicate
that the high density approximation is less accurate for $\surd
s=200$~GeV $pp$ collisions. Approximate KNO scaling has been
predicted to persist also for min-bias $p+Pb$ collisions (at LHC
energies) in spite of additional Glauber fluctuations of the number of
participants and binary collisions~\cite{Dumitru:2012yr}.
A more detailed discussion of multiplicity distributions at TeV
energies is given in refs.~\cite{Ugoccioni:2004wn}, and references therein.

Transverse momentum integrated multiplicities in inelastic hadronic
collisions are not governed by an external hard scale, unlike say
multiplicity distributions in $e^+ e^-$ annihilation or in
jets~\cite{KNOhard}. Hence, the
explanation for the experimental observation should relate to
properties of the distribution of produced gluons around the
saturation scale $Q_s$.

\section{KNO scaling from a Gaussian action in the classical limit}

We shall first discuss the multiplicity distribution of small-$x$
gluons obtained from a Gaussian effective action for the color charge
fluctuations of the valence charge densities $\rho$~\cite{MV},
\bea 
Z &=& \int {\cal D}\rho\; e^{-S_{MV}[\rho]}~,\\
S_{MV}[\rho] &=& \int d^2\bold x_\perp\int^\infty_{-\infty} dx^-~
\frac{\rho^a(x^-,\bold x_\perp) \, \rho^a(x^-,\bold
  x_\perp)}{2\mu^2(x^-)}~. \label{eq:S_MV}
\eea
In the strong field limit a semi-classical approximation is
appropriate and the soft gluon field (in covariant gauge) can be
obtained in the Weizs\"acker-Williams approximation as
\be
A^+(z^-,\bold x_\perp) = - g\, \frac{1}{\nabla^2_\perp}
\rho^a(z^-,\bold x_\perp) =  g \int d^2\bold
z_\perp \int\frac{d^2 \bold k_\perp}{(2\pi)^2}\frac{e^{i\bold k_\perp \cdot
    (\bold x_\perp-\bold z_\perp)}} {\bold k^2_\perp} \rho^a(z^-,\bold z_\perp)
~.
\ee
Parametrically, the mean multiplicity obtained from the
action~(\ref{eq:S_MV}) is then
\be \label{eq:nbar}
\bar{n} \sim \frac{N_c\, (N_c^2-1)}{\alpha_s} \; Q_s^2 S_\perp~,
\ee
where $S_\perp$ denotes a transverse area and $Q_s\sim g^2\mu$. The
prefactor in~(\ref{eq:nbar}) can be determined
numerically~\cite{KNV,ktF} but is not required for our present
considerations.

One can similarly calculate the probability to produce $q$ particles by
considering fully connected diagrams with $q$ valence sources $\rho$
in the amplitude and $q$ sources $\rho^*$ in the conjugate amplitude
(for both projectile and target, respectively). These can be expressed
as~\cite{Gelis:2009wh} \footnote{The rapidities $y_1\cdots y_q$ of the
$q$ particles should be similar. Here we assume that all particles are
in the same rapidity bin.}
\be
\left < \frac{d^q N}{dy_1\cdots dy_q} \right>_{\rm conn.} = C_q\,
\left < \frac{d N}{dy_1} \right> \cdots
\left < \frac{d N}{dy_q} \right>~,
\ee
where the reduced moments
\be \label{eq:C_q_MV}
C_q = \frac{(q-1)!}{k^{q-1}}~.
\ee
This expression is valid with logarithmic accuracy and was derived
under the assumption that all transverse momentum integrals over
$p_{T,1} \cdots p_{T,q}$ are effectively cut off in the infrared at a
scale $\sim Q_s$ due to non-linear effects.

The fluctuation parameter $k$ in eq.~(\ref{eq:C_q_MV}) is of order
\be
k \sim (N_c^2-1)\, Q_s^2\, S_\perp~.
\ee
Once again, the precise numerical prefactor (in the classical
approximation) has been determined by a numerical computation to all
orders in the valence charge density $\rho$~\cite{Lappi:2009xa}.

The multiplicity distribution is therefore a negative binomial
distribution (NBD)~\cite{Gelis:2009wh},
\be \label{eq:NBD}
P(n) = \frac{\Gamma(k+n)}{\Gamma(k) \, \Gamma(n+1)}
\frac{\bar n^n k^k}{(\bar n + k)^{n+k}}~.
\ee
Indeed, multiplicity distributions observed in high-energy $pp$
collisions (in the central region) can be described quite well by a
NBD, see for example refs.~\cite{Dumitru:2012yr,Tribedy:2010ab}. The
parameter $k^{-1}$ determines the variance of the
distribution\footnote{More precisely, the width is given by $\bar{n}\,
  \sqrt{k^{-1}+\bar{n}^{-1}}\sim \bar{n}/\surd k$; the latter
  approximation applies in the limit $\bar{n}/k\gg1$, see below.} and
can be obtained from the (inclusive) double-gluon multiplicity:
\be \label{eq:dN2_k}
\left < \frac{d^2 N}{dy_1 dy_2} \right>_{\rm conn.} = \frac{1}{k}\;
\left < \frac{d N}{dy_1} \right> \;
\left < \frac{d N}{dy_2} \right>~.
\ee
From this expression it is straightforward to see that the
perturbative expansion of $k^{-1}$ starts at ${\cal O}(\alpha_s^0)$
since the connected diagrams on the lhs of eq.~(\ref{eq:dN2_k})
involve the same number of sources and vertices as the disconnected
diagrams on the rhs of that equation (also see appendix). This
observation is important since {\em in general} the NBD~(\ref{eq:NBD})
exhibits KNO scaling only when $\bar{n}/k\gg1$, and if $k$ is not
strongly energy dependent. A numerical analysis of the multiplicity
distribution at 2360~GeV, for example, achieves a good fit to the data
for $\bar{n}/k\simeq 6-7$~\cite{Dumitru:2012yr}, which we confirm
below. Such values for $\bar{n}/k$ have also been found for peripheral
collisions of heavy ions from {\rm ab initio} solutions of the
classical Yang-Mills equations~\cite{Schenke:2012hg}; furthermore
those solutions predict that $\bar{n}/k<1$ for central collisions of
$A\sim200$ nuclei.

To illustrate how deviations from KNO scaling arise it is instructive
to consider a ``deformed'' theory with an additional contribution to
the quadratic action. We shall add a quartic operator~\cite{quartic},
\bea
S_Q[\rho] = \int d^2\bold v_\perp\int^\infty_{-\infty} dv_1^- \left\{
\frac{\rho^a(v_1^-,\bold v_\perp) \rho^a(v_1^-,\bold
  v_\perp)}{2\tilde\mu^2(v_1^-)}
+\int^\infty_{-\infty}dv_2^-\, \frac{\rho^a(v_1^-,\bold
  v_\perp)\rho^a(v_1^-,\bold v_\perp)\rho^b(v_2^-,\bold
  v_\perp)\rho^b(v_2^-,\bold v_\perp)}{\kappa_4}\right\}~.
\label{eq:Squartic}
\eea
We assume that the contribution from the quartic operator is a small
perturbation since $\kappa_4\sim g\, (gA^{1/3})^3$ while $\tilde\mu^2
\sim g\, (gA^{1/3})$. In the classical approximation the mean
multiplicity is unaffected by the correction as it involves only
two-point functions\footnote{The two-point functions
  $\langle\rho\rho\rangle$ in the theories~(\ref{eq:S_MV})
  and~(\ref{eq:Squartic}) need to be matched. Thus, the ``bare'' parameters
  $\mu^2$ in~(\ref{eq:S_MV}) and $\tilde\mu^2$ in~(\ref{eq:Squartic})
  are {\em different} as the latter absorbs some self-energy
  corrections. We refer to ref.~\cite{quartic} for
  details.}.  On the other hand, $k^{-1}$ as defined
in~(\ref{eq:dN2_k}) now becomes
\bea \label{eq:k_quartic}
\frac{N_c^2-1}{2\pi} Q_s^2\, S_\perp \frac{1}{k} &=&  1 - 3\beta
\, (N_c^2+1)  ~~~~~~~~\left(\mathrm{with}~
\beta \equiv \frac{C_F^2}{6\pi^3}\frac{g^8}{Q_s^2\, \kappa_4} 
\left[\int_{-\infty}^\infty dz^- \mu^4(z^-)\right]^2 \right)~.
\eea
\begin{figure}[htb]
\begin{center}
\includegraphics[width=0.47\textwidth]{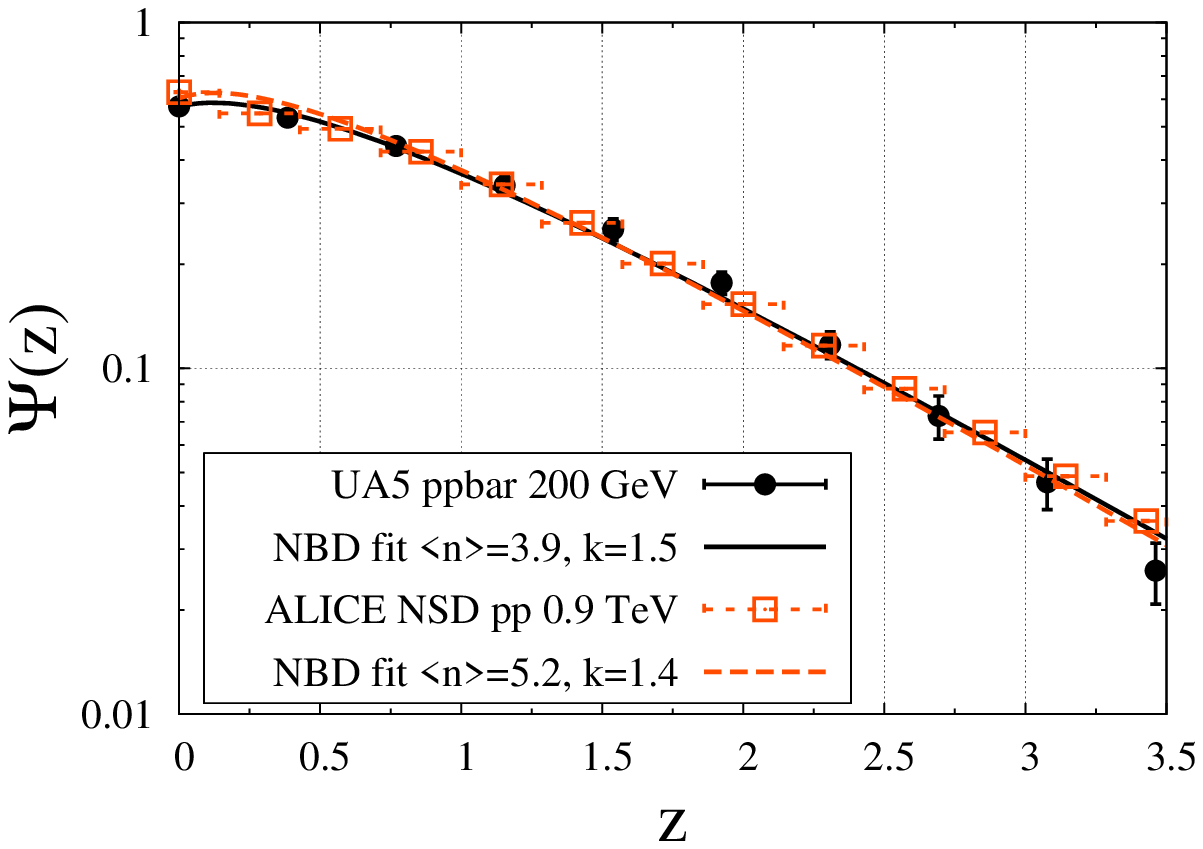}
\includegraphics[width=0.47\textwidth]{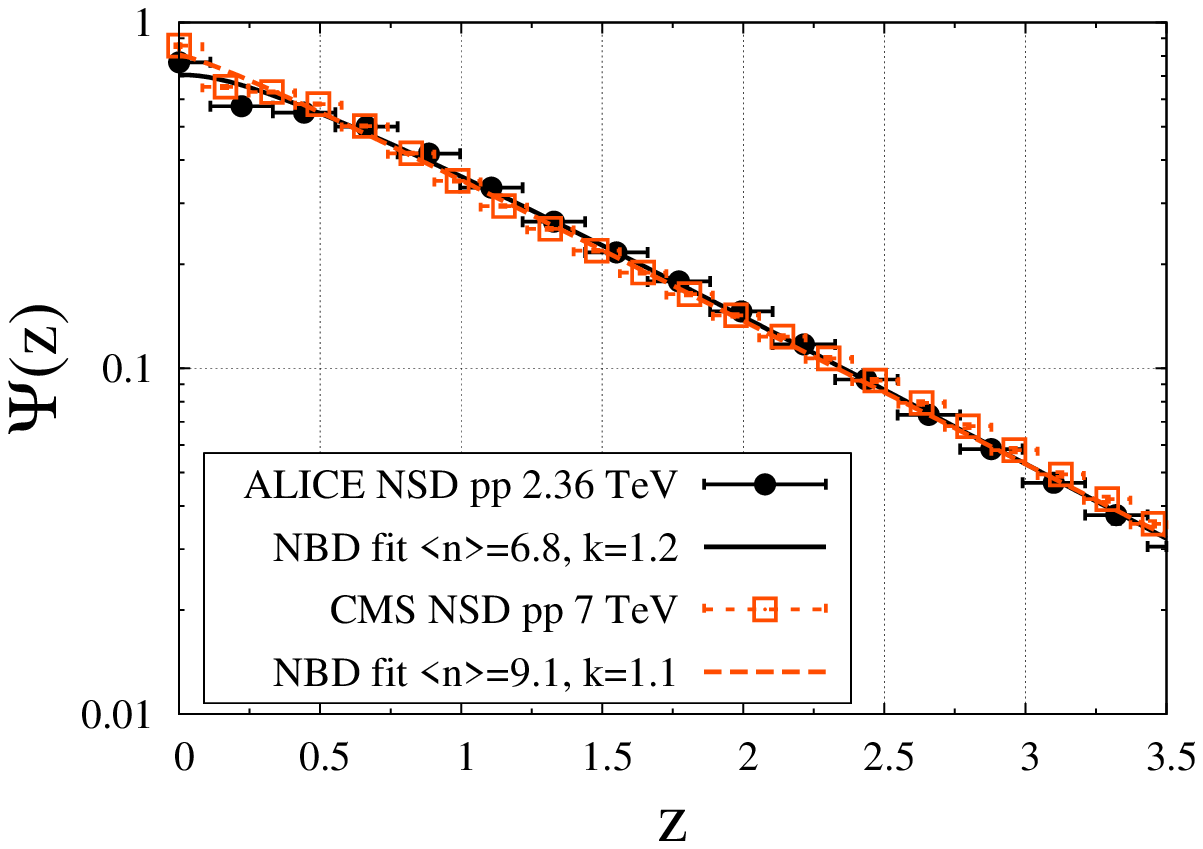}
\end{center}
\caption[a]{KNO scaling plots of charged particle multiplicity
  distributions at $|\eta|<0.5$ in NSD collisions at various energies
  and NBD fits; $z\equiv N_{\rm ch}/\langle N_{\rm ch}\rangle$ and
  $\Psi(z)\equiv \langle N_{\rm ch}\rangle\, P(N_{\rm ch})$. Note
  that the mean multiplicity quoted for the fits has been rescaled by
  1.5 to include neutral particles; also, that here $k$ is integrated
  over the transverse plane of the collision.}
\label{fig:KNOfits}
\end{figure}
Therefore, in the classical approximation
\be \label{eq:Quartic_nbar_k}
\frac{\bar{n}}{k} \sim \frac{N_c}{\alpha_s} \left( 1 - 3\beta\,
(N_c^2+1) \right)~.
\ee
This result illustrates that $\bar{n}/k$ decreases as the contribution
of the $\sim\rho^4$ operator increases. We repeat that the derivation assumed
that the correction is small so that~(\ref{eq:Quartic_nbar_k}) does
not apply for large values of $\beta N_c^2$.

Ref.~\cite{quartic} estimated by entirely different considerations
that for protons $\beta\simeq0.01$ at $x=10^{-2}$. That would
correspond to a smaller value of $\bar{n}/k$ by a factor of 1.43 than
for the Gaussian theory. Assuming that RG flow with energy approaches
a Gaussian action~\cite{Iancu:2002aq}, $\bar{n}/k$ should increase by
about this factor. NBD fits to the data shown in
fig.~\ref{fig:KNOfits} confirm that $\bar{n}/k$ indeed increases with
energy, which might indicate flow towards a Gaussian action; however,
the observed increase from $\surd s=200$~GeV to 7~TeV is much
stronger: a factor of about 3. This apparent discrepancy could be
resolved at least partially by running of the coupling in
eqs.~(\ref{eq:nbar},\ref{eq:Quartic_nbar_k}) with $Q_s$ but this
requires more careful analysis\footnote{In fact, the running of
  $\alpha_s$ at the effective scale $Q_s$ is taken into account if the
  mean multiplicity is computed with energy evolved unintegrated gluon
  distributions like e.g.\ in refs.~\cite{Dumitru:2012yr,ktF}.}.

\section{Quantum evolution and the distribution of dipoles in the
  hadronic wave function}

In the previous section we considered the multiplicity distribution of
``produced gluons'' in a collision of classical YM fields sourced by
classical color charges $\rho$ moving on the light-cone. At high
energies though (i.e., when $\alpha_s\log x^{-1}\sim1$) the classical
fields are modified by quantum
fluctuations~\cite{Mueller:1999wm}. Resummation of boost-invariant
quantum fluctuations leads to an energy dependent saturation scale,
for example, as required in order to reproduce the growth of the
multiplicity $\bar{n}$ with energy. In particular, the energy
dependence of the {\em mean} saturation scale, averaged over all
``evolution ladders'' (distribution of quantum emissions), can be obtained
by solving the running-coupling BK equation~\cite{rcBK}.

Instead, in this section we shall solve a {\em stochastic} evolution
equation which accounts both for saturation (non-linear) effects as
well as for the fluctuations of the rapidities and transverse momenta
of the virtual gluons in the wave function of a hadron before the
collision. We do this in order to determine the multiplicity
distribution (rather than just the mean number) of dipoles in a
hadronic wave function boosted to rapidity $Y$.

We shall do so by solving via Monte-Carlo techniques the following
evolution equation for $P[n(x),Y]$, which is the probability for the
dipole size distribution $n(x)$ to occur:
\be \label{eq:PLevol}
\frac{\partial P[n(x),Y]}{\partial Y} = \int\limits_z
 f_z[n(x)-\delta_{xz}]\,P[n(x)-\delta_{xz},Y] -\int\limits_z
 f_z[n(x)]\,P[n(x),Y]~.
\ee
Note that in this section $x=\log\,1/r^2$ denotes the logarithmic
dipole size (conjugate to its transverse momentum) rather than to a
light-cone momentum fraction. This equation has been studied before in
ref.~\cite{Iancu:2006jw} for fixed $\alpha_s$ and in
ref.~\cite{Dumitru:2007ew} for running $\alpha_s(r^2)$. Those papers
also provide references to related earlier work.

The first term in~(\ref{eq:PLevol}) is a gain term due to dipole
splitting while the second term corresponds to loss due to
``recombination''.
\be
f_z[n(x)] = \frac{T_z[n(x)]}{\alpha(z)}
\ee
is the splitting rate and
\be
T_z[n(x)] = 1 - \exp
\int \limits_x n(x) \;\log\left(1-\tau(z|x)\right)
\ee
is the dipole scattering amplitude for a dipole projectile of size $z$
to scatter off the target with the dipole distribution $n(x)$. Note
that $T_z[n(x)]$ is non-linear in the dipole density as it involves
also the {\em pair} (and higher) densities. Finally,
\be
\tau(x|y) = \alpha(x) \alpha(y) \exp(-|x-y|) 
\equiv
 \alpha(r^2_<) \,\alpha(r^2_>) \, \frac{r^2_<}{r^2_>}
\ee
is the elementary dipole-dipole scattering amplitude at LO in
perturbative QCD. For more details we refer to
ref.~\cite{Dumitru:2007ew}. Here, we recall only that it was found
there that evolution with a running coupling suppresses fluctuations
in the tails of the travelling waves and so restores approximate
geometric scaling~\cite{Stasto:2000er}.

We have determined the multiplicity distribution of dipoles with size
$\sim 1/Q_s(Y)$,
\be
N_i(Y) = \int\limits_{r^2 < 1/\langle Q_s^2(Y)\rangle} dx ~ n_i(x,Y)
\hspace{2cm} (i=1\cdots 10^5)
\ee
by evolving a given initial configuration $n(x,Y=0)$ $10^5$
times. Despite starting with a fixed initial condition, evolution
introduces fluctuations in the rapidities where splittings occur, and
in the sizes of the emerging dipoles.

\begin{figure}[htb]
\begin{center}
\includegraphics[width=0.47\textwidth]{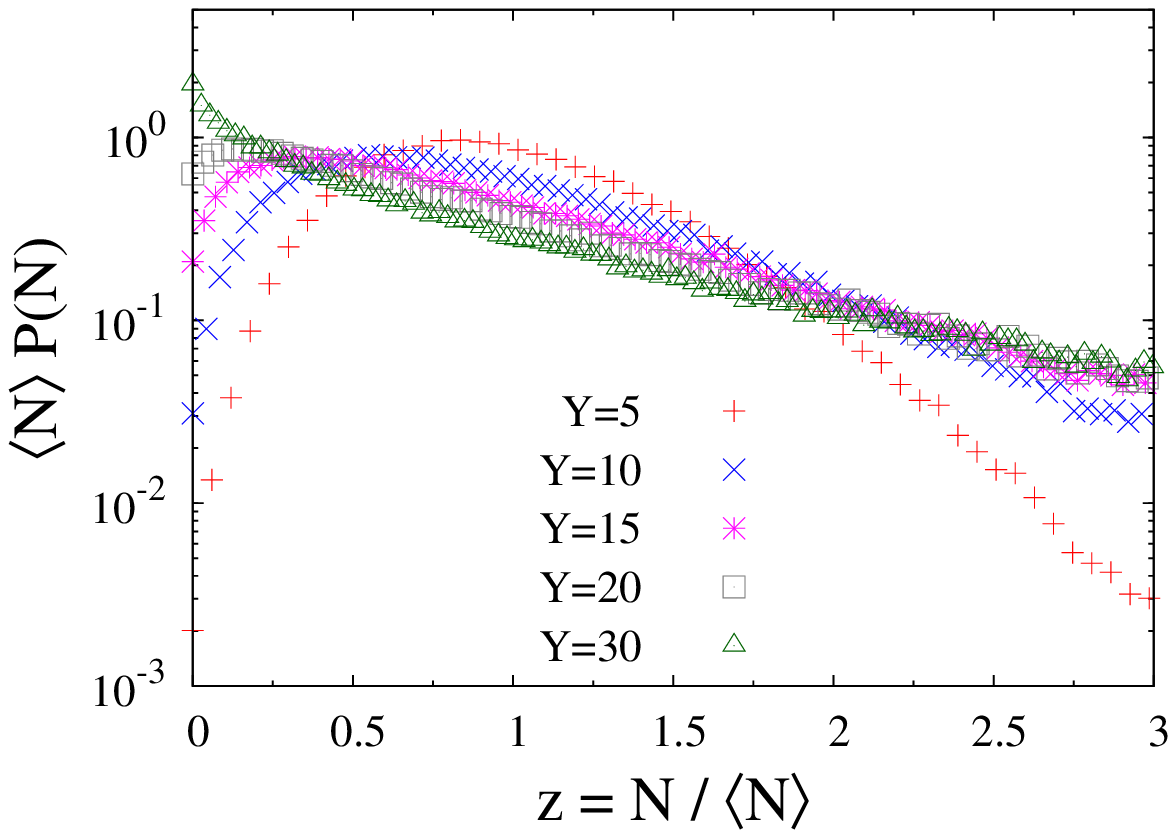}
\includegraphics[width=0.47\textwidth]{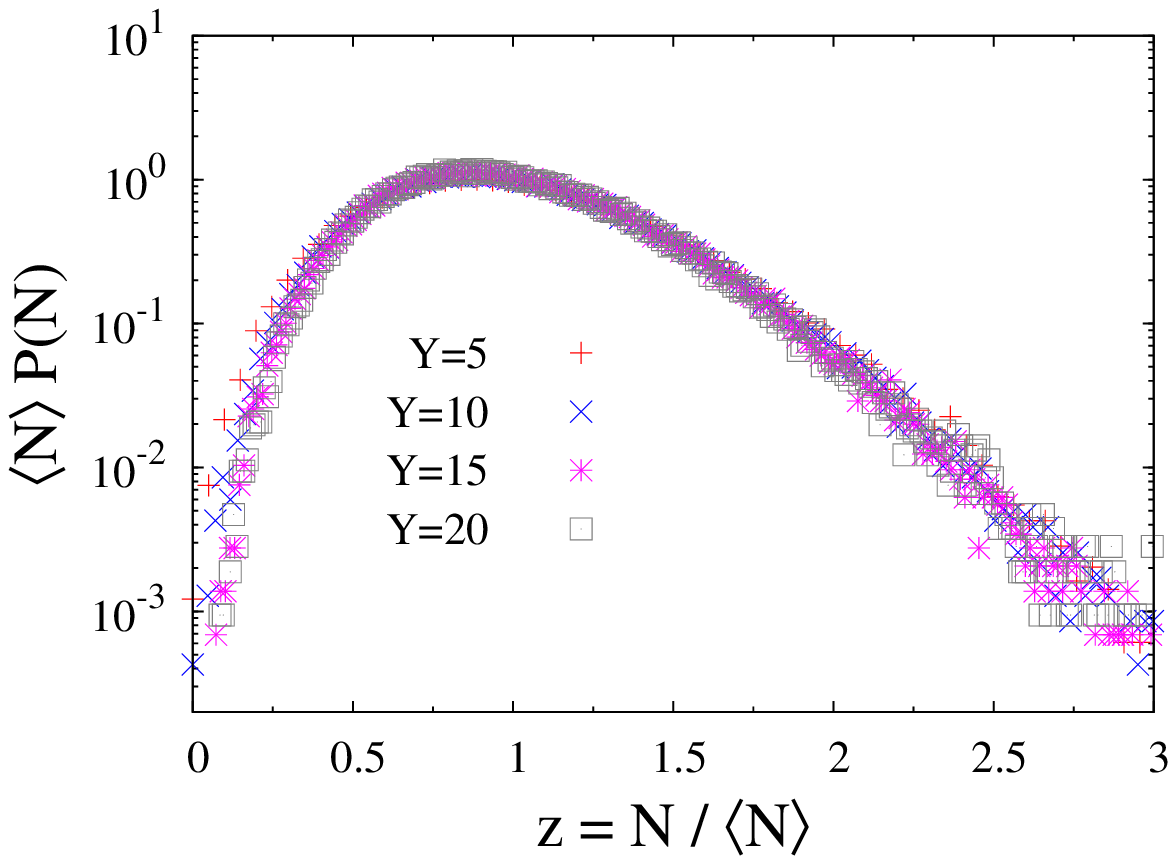}
\end{center}
\caption[a]{Multiplicity distribution of virtual dipoles with size
  $r^2 < 1/\langle Q_s^2(Y)\rangle$ in the wave function. Left:
  evolution with trivial $\beta$-function, $\alpha_s=$const. Right:
  QCD $\beta$-function.}
\label{fig:PL_KNO}
\end{figure}
In fig.~\ref{fig:PL_KNO} we show that fixed-coupling evolution does
not obey KNO scaling of the distribution of virtual quanta while
running-coupling evolution does. The shape of the distribution however
looks different than the measured distribution of produced particles
from fig.~\ref{fig:KNO_LHCdata}. This could be due to the fact that
our evolution model does not treat diffusion in impact parameter
space. Hence, $P(N)$ shown in fig.~\ref{fig:PL_KNO} should be
interpreted as the multiplicity distribution at the center of the hadron.

\begin{acknowledgments}
Our work is supported by the DOE Office of Nuclear Physics through
Grant No.\ DE-FG02-09ER41620 and by The City University of New York
through the PSC-CUNY Research Award Program, grant 65041-0043.
\end{acknowledgments}

\appendix
\section{The moment $C_2 = k^{-1}$ with the quartic action}
We can obtain the fluctuation parameter $k$ by calculating the inclusive
double gluon multiplicity and expressing it in terms of the single
inclusive or mean multiplicity. The connected two particle production
cross section for gluons with rapidity $y_1$ and $y_2$ has the form:
\be
N_2(p,q) \equiv \left < \frac{d^2N}{dy_1 dy_2} \right > -\left <
\frac{dN}{dy_1} \right > \left < \frac{dN}{dy_2} \right > \equiv \left
< \frac{d^2N}{dy_1 dy_2} \right > _{conn.} \label{eq:C_2}
\ee
$ \left < \frac{dN}{dy} \right > $ is the mean multiplicity and the
brackets denote an average over events. $N_2( p,  q)$ is given by:
\bea
N_2(p, q)=\frac{g^{12}}{4(2 \pi)^6}f_{gaa'}f_{g'bb'}f_{gcc'}f_{g'dd'} \int \prod\limits_{i=1}^4 \frac{d^2k_i}{(2\pi)^2k^2_i} \frac{L_\mu(p,k_1)L^\mu(p,k_2)L_\nu(q,k_3)L^\nu(q,k_4)}{(p-k_1)^2(p-k_2)^2(q-k_3)^2(q-k_4)^2} \times \nonumber \\
\langle \rho_1^{*a}(k_2) \rho_1^{*b}(k_4) \rho_1^{c}(k_1)
\rho_1^{d}(k_3) \rangle \langle \rho_2^{*a'}(p-k_2)
\rho_2^{*b'}(q-k_4) \rho_2^{c'}(p-k_1) \rho_2^{d'}(q-k_3)
\rangle~.~~~~~ \nonumber
\eea
$L^\mu$ denotes the Lipatov vertex, for which:
\be
L_\mu(p,k)L^\mu(p,k)=-\frac{4k^2}{p^2}(p-k)^2 ~.\label{eq:LipatovVertex}
\ee

For the four-point function in the target and projectile fields we
use~\cite{quartic}
\bea
\langle \rho^{*a'}_2(p-k_2) \rho^{*b'}_2(q-k_4) \rho^{c'}_2(p-k_1) \rho^{d'}_2(q-k_3) \rangle = ~~~~~~~~~~~~~~~~~~~~~~~~~~~~~~~~~~~~~~~~~~~~~
 \nonumber\\
 (2\pi)^4\, \left [\int dz^- \tilde\mu^2(z^-)\right ]^2 \left[
  \delta^{a'b'}\delta^{c'd'}\delta(p+q-k_2-k_4)\delta(p+q-k_1-k_3)\right .~~~~~~~~~~~~\nonumber \\
 \left .  +
  \delta^{a'c'}\delta^{b'd'}\delta(k_1-k_2)\delta(k_3-k_4) +
  \delta^{a'd'}\delta^{b'c'}\delta(p-q-k_2+k_3)\delta(p-q-k_1+k_4)\right ]
     \nonumber\\
 -(2\pi)^4  \frac{2}{\pi^2\kappa_4} \left[ \int dz^-\tilde\mu^4(z^-)\right]^2
    \left(\delta^{a'b'}\delta^{c'd'} + \delta^{a'c'}\delta^{b'd'} + \delta^{a'd'}\delta^{b'c'}
    \right)
    \delta(k_1+k_3-k_2-k_4)~. \nonumber \label{eq:4ptCorr}
\eea
The first two lines on the rhs of the above equation originate from
the quadratic part of the action while the third line is due to the
quartic operator. The product of the Gaussian parts of the two
four-point functions gives nine terms, one of which $(\sim
\delta^{ac}\delta^{bd}\delta^{a'c'}\delta^{b'd'})$ corresponds to a
disconnected contribution. It exactly cancels the second term in
eq.~(\ref{eq:C_2}).

\begin{figure}[htb]
\includegraphics[width=50mm]{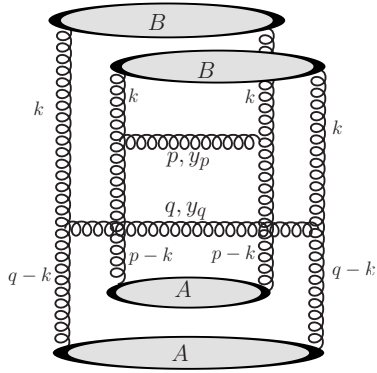}
\caption{One of eight connected diagrams for two-gluon production with
  the quadratic MV action.}
\label{fig:GaussianConn}
\end{figure}
\begin{figure}[htb]
\includegraphics[width=34mm]{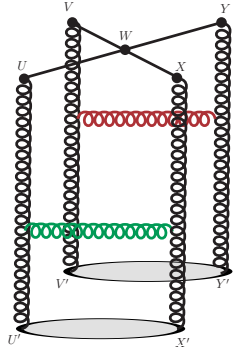}
\caption{Connected diagram for two-gluon production from the quartic
  operator in the action~\cite{quartic}.}
\label{fig:QuarticConn}
\end{figure}

Four of the other eight terms $(\sim \delta^{ac}\delta^{bd}$ or $\sim
\delta^{a'c'}\delta^{b'd'})$ give identical leading contributions to
double gluon production. They correspond to a ``rainbow'' diagram like
the one shown in Fig.~\ref{fig:GaussianConn}. In the ``rainbow''
diagram, on one side (target or projectile), the $\rho$'s
corresponding to the same gluon momentum are contracted with each
other. The remaining four ``non-rainbow'' diagrams are suppressed
relative to the terms we keep at large $p$ and
$q$~\cite{Gelis:2009wh}. Hence, the leading Gaussian contribution is:
\be
\sim \frac{g^{12}}{8\pi^7}\left [\int dz^- \tilde\mu^2(z^-)\right ]^4
\frac{S_\perp}{Q_s^2} \frac{N_c^2(N_c^2-1)}{p^4q^4}~. \label{eq:GaussianConn}
\ee
The same reasoning applies also for the additional quartic
contribution and only ``rainbow'' diagrams are considered, like the one
in Fig.~\ref{fig:QuarticConn}. There are two of them (one for the
projectile and one for the target) to first order in $\kappa_4^{-1}$,
and their contribution is:
\bea \sim -\frac{g^{12}}{2(2 \pi)^6}f_{gaa'}f_{g'bb'}f_{gcc'}f_{g'dd'}
\int \prod\limits_{i=1}^4 \frac{d^2k_i}{(2\pi)^2k^2_i}
\frac{L_\mu(p,k_1)L^\mu(p,k_2)L_\nu(q,k_3)
  L^\nu(q,k_4)}{(p-k_1)^2(p-k_2)^2(q-k_3)^2(q-k_4)^2}
\times \nonumber \\ \frac{2(2 \pi)^8}{\pi^2 \kappa_4} \left [\int dz^-
  \tilde\mu^2(z^-)\right ]^2 \left [\int dz^- \tilde\mu^4(z^-)\right
]^2 \delta^{ac}\delta^{bd}\delta(k_1-k_2)\delta(k_3-k_4)\times
\nonumber
\\ \left(\delta^{a'b'}\delta^{c'd'} + \delta^{a'c'}\delta^{b'd'} +
\delta^{a'd'}\delta^{b'c'}\right)\delta(k_1+k_3-k_2-k_4)
~.\nonumber 
\eea
The color factor evaluates to
\be f_{gaa'}f_{g'bb'}f_{gcc'}f_{g'dd'} \delta^{ac}\delta^{bd}
\left(\delta^{a'b'}\delta^{c'd'}+\delta^{a'c'}\delta^{b'd'}+
 \delta^{a'd'}\delta^{b'c'}\right)=2 N_c^2(N_c^2-1)+N_c^2(N_c^2-1)^2
~.\nonumber \ee
Using eq.~(\ref{eq:LipatovVertex}) we get:
\bea -\frac{16 g^{12}}{(2\pi)^8\pi^2\kappa_4}\left [\int dz^-
  \tilde\mu^2(z^-)\right ]^2 \left [\int dz^- \tilde\mu^4(z^-)\right
]^2\left[2 N_c^2(N_c^2-1)+N_c^2(N_c^2-1)^2\right]\times \nonumber
\\ \frac{S_\perp}{p^2q^2}\int \frac{d^2k_1}{k_1^2(p-k_1)^2} \int
\frac{d^2k_3}{k_3^2(q-k_3)^2} ~.\nonumber 
\eea
The integral over the ladder momentum is again cut off at the
saturation scale $Q_s$:
\be
\int \frac{d^2k_1}{k_1^2(p-k_1)^2}\approx
\frac{2\pi}{p^2}\log\frac{p}{Q_s}~. \nonumber
\ee
Then, the quartic contribution to connected two gluon production
becomes
\be
 - \frac{g^{12}}{4 \pi^8 \kappa_4}\left [\int dz^-
   \tilde\mu^2(z^-)\right ]^2 \left [\int dz^- \tilde\mu^4(z^-)\right
 ]^2 \left[2
   N_c^2(N_c^2-1)+N_c^2(N_c^2-1)^2\right]\frac{S_\perp}{p^4q^4} \log
 \frac{p}{Q_s}\log
 \frac{q}{Q_s}~. \label{eq:QuarticConn}
\ee
The last step is to express the fully connected diagrams in terms of
the single inclusive cross section:
\be
\left < \frac{dN}{dy} \right > = \frac{g^6}{4\pi^4}\left [\int dz^-
  \tilde\mu^2(z^-)\right ]^2
N_c(N_c^2-1)\frac{S_\perp}{p^4}\log\frac{p}{Q_s} ~. \label{eq:Single}
\ee
Summing eq.~(\ref{eq:GaussianConn}) and eq.~(\ref{eq:QuarticConn}) and
using eq.~(\ref{eq:Single}) we get:
\be
\left < \frac{d^2N}{dy_1 dy_2} \right >
_{conn.}=\left[\frac{2\pi}{Q_s^2(N_c^2-1)S_\perp} - \frac{4(N_c^2+1)}
  {\kappa_4S_\perp(N_c^2-1)}
  \frac{\left
    [\int dz^- \tilde\mu^4(z^-)\right ]^2}{\left [\int dz^-
      \tilde\mu^2(z^-)\right ]^2}\right]\left < \frac{dN}{dy_1} \right
> \left < \frac{dN}{dy_2} \right > ~.\nonumber 
\ee
The fluctuation parameter $k^{-1}$ is now identified with the
expression in the square brackets. We rewrite it in terms of
\be \label{eq:beta}
\beta \equiv \frac{C_F^2}{6\pi^3}\frac{g^8}{Q_s^2\kappa_4} \left [\int
  dz^- \tilde\mu^4(z^-)\right ]^2~,
\ee
and use
\be \label{eq:Qs}
Q_s^2=\frac{g^4C_F}{2\pi}\int dz^- \tilde\mu^2(z^-)~,
\ee
to arrive at the final expression
\be  \label{eq:1/k}
\frac{N_c^2-1}{2\pi}\, Q_s^2S_\perp\, \frac{1}{k} = 1 - 3\beta
(N_c^2+1)~.
\ee

\section{The moment $C_3$ with the quartic action}

In this section we are going to calculate the connected diagrams for
three-gluon production to obtain the correction to the reduced moment
$C_3$ at order $1/\kappa_4 \sim 1/[g(gA^{1/3})^3]$, assuming as before
that $gA^{1/3}>1$. At the end of this section we also outline
corrections suppressed by higher powers of $gA^{1/3}$.

We are looking for the contribution of the connected diagrams to the
following expression~\cite{DuslingFernandezVenugopalan}:
\bea \left < \frac{d^3N}{dy_1 dy_2 dy_3} \right > =
\frac{(-ig^3)^6}{8(2\pi)^9}f_{gaa'}f_{g'bb'}f_{g''cc'}f_{gff'}f_{g'ee'}f_{g''dd'}\times
~~~~~~~~~~~~~~~~~~~~~~~~~~~~~~~~~~~\nonumber \\ \int
\prod\limits_{i=1}^6 \frac{d^2k_i}{(2\pi)^2k^2_i}
\frac{L_\alpha(p,k_1)L^\alpha(p,k_2)L_\beta(q,k_3)L^\beta(q,k_4)
  L_\gamma(l,k_5) L^\gamma(l,k_6)}
     {(p-k_1)^2
  (p-k_2)^2
  (q-k_3)^2
  (q-k_4)^2
  (l-k_5)^2
  (l-k_6)^2}
\times \nonumber \\ \langle \rho_1^{*f}(p-k_2) \rho_1^{*e}(q-k_4)
\rho_1^{*d}(l-k_6) \rho_1^{a}(p-k_1)\rho_1^{b}(q-k_3)
\rho_1^{c}(l-k_5) \rangle \times \nonumber \\ \langle
\rho_2^{*f'}(k_2) \rho_2^{*e'}(k_4) \rho_2^{*d'}(k_6)
\rho_2^{a'}(k_1)\rho_2^{b'}(k_3) \rho_2^{c'}(k_5)
\rangle~.~~~~~~~~~~~~~~~~~~~~~~~~~~~~~ \label{eq:C3} 
\eea
As before, the $\rho$ correlators of the target and the projectile consist
of two parts, one from the quadratic operator in the action and
another from the additional $\rho^4$ operator:
\be
\langle \rho^{*f} \rho^{*e} \rho^{*d} \rho^{a}\rho^{b} \rho^{c}
\rangle = \langle \rho^{*f} \rho^{*e} \rho^{*d} \rho^{a}\rho^{b}
\rho^{c} \rangle _{\text{Gaussian}} + \langle \rho^{*f} \rho^{*e}
\rho^{*d} \rho^{a} \rho^{b} \rho^{c} \rangle
_{\text{Correction}}~. \nonumber
\ee
The product of the two Gaussian contributions from the target and the
projectile, to leading order in the gluon momenta, gives rise to 16
"rainbow" diagrams. The result has been obtained
previously~\cite{Gelis:2009wh} and reads (expressed in terms of the
mean multiplicity):
\be
\left < \frac{d^3N}{dy_1 dy_2 dy_3} \right >_{\text{Conn. Gaussian}} =
\frac{8\pi^2}{Q_s^4S_\perp^2(N_c^2-1)^2} \left < \frac{dN}{dy_1}
\right > \left < \frac{dN}{dy_2} \right > \left < \frac{dN}{dy_3}
\right >~. \label{eq:Gaussian3Gluon}
\ee
The correction, to first order in $\kappa_4^{-1}$ is
\be
\sim 2 \langle \rho^{*f'} \rho^{*e'} \rho^{*d'} \rho^{a'}\rho^{b'} \rho^{c'} \rangle _{\text{Gaussian}} ~\langle \rho^{*f} \rho^{*e} \rho^{*d} \rho^{a} \rho^{b} \rho^{c} \rangle _{\text{Correction}}~. \label{eq:simCorr}
\ee
Again, we are considering only rainbow diagrams, so for the Gaussian
six-point function in the above expression, from all possible
contractions, we keep only the term
\be
(2\pi)^6 \left [\int dz^- \tilde\mu^2(z^-)\right ]^3 \delta^{a'f'}\delta^{b'e'}\delta^{c'd'}\delta(k_1-k_2)\delta(k_3-k_4)\delta(k_5-k_6)~. \nonumber
\ee
To calculate the correction to the six-point function to first order
in $\kappa_4^{-1}$ we factorize it into a product of two- and four-point
functions. There are fifteen possible factorizations of that
kind. Three of them are disconnected diagrams and the remaining twelve
give identical contributions.  We consider, for example, the following
combination:
\bea
\langle \rho_1^{*f}(p-k_2) \rho_1^{*e}(q-k_4) \rho_1^{*d}(l-k_6)
\rho_1^{a}(p-k_1)\rho_1^{b}(q-k_3) \rho_1^{c}(l-k_5)
\rangle~~~~~~~~~~~~~~~~~~~~~~~~~~~~~~~~~~ \nonumber \\
 = \langle \rho_1^{a}(p-k_1)\rho_1^{b}(q-k_3) \rangle \langle \rho_1^{*f}(p-k_2) \rho_1^{*e}(q-k_4) \rho_1^{*d}(l-k_6) \rho_1^{c}(l-k_5) \rangle~.~~~~~~~~~~~~~~~~~~~~~ \nonumber
\eea
The two point function is
\be
\langle \rho_1^{a}(p-k_1)\rho_1^{b}(q-k_3) \rangle = (2\pi)^2 \left [ \int dz^- \tilde\mu^2(z^-)\right ] \delta^{ab} \delta(p+q-k_1-k_3)~, \nonumber
\ee
and for the correction to the four-point function we use the last line
from eq.~(\ref{eq:4ptCorr}).

The color factor is
\bea
f_{gaa'}f_{g'bb'}f_{g''cc'}f_{gff'}f_{g'ee'}f_{g''dd'}\delta^{ab}\left(\delta^{cd}\delta^{ef}
+ \delta^{ce}\delta^{df} + \delta^{cf}\delta^{de} \right)
\delta^{a'f'}\delta^{b'e'}\delta^{c'd'} \nonumber \\
=2N_c^3(N_c^2-1)+N_c^3(N_c^2-1)^2~.~~~~~~~~~~~~~~~~~~~~~~~~~~~~ \nonumber
\eea
Putting everything together into eq.~(\ref{eq:C3}) and multiplying by two
[because of (\ref{eq:simCorr})] and by twelve (which is the number of possible
diagrams) we get:
\bea
\left < \frac{d^3N}{dy_1 dy_2 dy_3} \right >_{\text{Conn. Correction}}
=\frac{48g^{18}\pi}{\kappa_4}\left [\int dz^- \tilde\mu^2(z^-)\right
]^4\left [\int dz^- \tilde\mu^4(z^-)\right ]^2
\left[2N_c^3(N_c^2-1)+N_c^3(N_c^2-1)^2\right] \times \nonumber \\
\int \prod\limits_{i=1}^6 \frac{d^2k_i}{(2\pi)^2k^2_i}
\frac{L_\alpha(p,k_1)L^\alpha(p,k_2)L_\beta(q,k_3)L^\beta(q,k_4)L_\gamma(l,k_5)L^\gamma(l,k_6)}{(p-k_1)^2(p-k_2)^2(q-k_3)^2(q-k_4)^2(l-k_5)^2(l-k_6)^2}
\times ~~~~~~~~ \nonumber \\ 
\delta(k_1-k_2)\delta(k_3-k_4)\delta(k_5-k_6)\delta(p+q-k_1-k_3)\delta(p+q-k_2-k_4-k_6+k_5)\nonumber
\\
\nonumber \\
=-\frac{3g^{18}}{16\kappa_4\pi^{13}}\frac{S_\perp}{p^2q^2l^2}\left
[\int dz^- \tilde\mu^2(z^-)\right ]^4\left [\int dz^-
  \tilde\mu^4(z^-)\right ]^2
\left[2N_c^3(N_c^2-1)+N_c^3(N_c^2-1)^2\right] \times \nonumber \\
\int\frac{d^2k_1}{k_1^2(p+q-k_1)^2(p-k_1)^4}\int\frac{d^2k_2}{k_2^2(l-k_2)^2}~. 
\nonumber
\eea
Again, we regularize the ladder integrals at the saturation scale,
\be
\int\frac{d^2k}{k^2(p+q-k)^2(p-k)^4}\simeq
\frac{2\pi}{p^2q^2}\frac{1}{Q_s^2}~. \nonumber
\ee
Finally, using expression (\ref{eq:Single}) for the mean multiplicity
the $\rho^4$ contribution to three-gluon production becomes
\be
\left < \frac{d^3N}{dy_1 dy_2 dy_3} \right
>_{\text{Conn. Correction}}=-\frac{48 \pi
  (N_c^2+1)}{\kappa_4Q_s^2S_\perp^2(N_c^2-1)^2}\frac{\left [\int dz^-
    \tilde\mu^4(z^-)\right ]^2}{\left [\int dz^-
    \tilde\mu^2(z^-)\right ]^2}\left < \frac{dN}{dy_1} \right >\left <
\frac{dN}{dy_2} \right >\left < \frac{dN}{dy_3} \right
>~. \label{eq:Correction3Gluon}
\ee
Summing (\ref{eq:Gaussian3Gluon}) and (\ref{eq:Correction3Gluon}),
\be
\left < \frac{d^3N}{dy_1 dy_2 dy_3} \right >_{\text{Conn.}} = \left[
  \frac{8\pi^2}{Q_s^4S_\perp^2(N_c^2-1)^2}-\frac{48 \pi
    (N_c^2+1)}{\kappa_4Q_s^2S_\perp^2(N_c^2-1)^2}\frac{\left [\int
      dz^- \tilde\mu^4(z^-)\right ]^2}{\left [\int dz^-
      \tilde\mu^2(z^-)\right ]^2}\right ]\left < \frac{dN}{dy_1}
\right >\left < \frac{dN}{dy_2} \right >\left < \frac{dN}{dy_3} \right
>~. \nonumber
\ee
From the above equation the third reduced moment is:
\be
C_3 =
\frac{8\pi^2}{Q_s^4S_\perp^2(N_c^2-1)^2}-\frac{48 \pi
  (N_c^2+1)}{\kappa_4Q_s^2S_\perp^2(N_c^2-1)^2}\frac{\left [\int dz^-
    \tilde\mu^4(z^-)\right ]^2}{\left [\int dz^-
    \tilde\mu^2(z^-)\right ]^2}~, \nonumber
\ee
or
\be \label{eq:C3final}
\frac{(N_c^2-1)^2}{4\pi^2}Q_s^4S_\perp^2\frac{C_3}{2}= 1 -
9\beta(N_c^2+1)~,
\ee
where we have used expressions (\ref{eq:beta}) and (\ref{eq:Qs}).

For a NBD we have that $C_3=2/k^2$ but if we compare~(\ref{eq:C3final})
to the square of eq.~(\ref{eq:1/k}), which is
\be
\frac{(N_c^2-1)^2}{4\pi^2}Q_s^4S_\perp^2\frac{1}{k^2}=1-6\beta(N_c^2+1)~,
\ee
we see that the coefficients of the corrections at order ${\cal
  O}(\beta)$ differ. That means that the $\rho^4$ operator
in the action provides a correction to the negative binomial
distribution.

In fact, such deviation from a NBD is more obvious if even higher
order operators are added to the action. Dropping the longitudinal
dependence of the operators for simplicity, such an action would have
the form
\bea
S\simeq \int d^2\bold v_\perp \left[ \frac {\delta^{ab} \rho^a
    \rho^b}{2\tilde
    \mu^2}-\frac{d^{abc}\rho^a\rho^b\rho^c}{\kappa_3}+\frac{\delta^{ab}\delta^{cd}+\text{perm.}}{\kappa_4}
  \rho^a\rho^b\rho^c\rho^d -
  \frac{\delta^{ab}d^{cde}+\text{perm.}}{\kappa_5}\rho^a\rho^b\rho^c\rho^d\rho^e
  \right. \nonumber \\
\left. +\frac{(\delta^{ab}\delta^{cd}\delta^{ef}+\text{perm.})+(d^{abc}d^{def}+\text{perm.})}{\kappa_6}
\rho^a\rho^b\rho^c\rho^d\rho^e\rho^f +\dots \right ]~. \nonumber
\eea
The additional terms are suppressed by powers of $gA^{1/3}$~\cite{rho_n_action}:
\bea
\tilde \mu^2\sim g(gA^{1/3})~,~~~~~~\kappa_3\sim
g(gA^{1/3})^2~,~~~~~~\kappa_4\sim g(gA^{1/3})^3~,~~~~~~
\kappa_5\sim g(gA^{1/3})^4~,~~~~~~\kappa_6\sim
g(gA^{1/3})^5~. \nonumber
\eea
The cubic operator gives a correction to the six-point function,
i.e.\ to $C_3$ at order $1/\kappa_3^2$ but does not correct the
four-point function, i.e. $C_2=1/k$ (it only renormalizes
$\mu^2$). The same applies to the $\rho^6$ operator: $C_3$ will
contain a term $\sim 1/\kappa_6$ but $1/k$ does not. Hence, beyond a
quadratic action the relation $C_3 = 2/k^2$ is not exact.



\begin{thebibliography}{99}

\bibitem{Mueller:1999wm} 
A.~H.~Mueller,
Nucl.\ Phys.\ B {\bf 558}, 285 (1999).

\bibitem{MV}
L.~D.~McLerran and R.~Venugopalan,
Phys.\ Rev.\ D49 (1994) 2233, Phys.\ Rev.\ D49 (1994) 3352;
Yu.~V.~Kovchegov, Phys.\ Rev.\ D54 (1996) 5463.

\bibitem{Koba:1972ng} 
Z.~Koba, H.~B.~Nielsen and P.~Olesen,
Nucl.\ Phys.\ B {\bf 40}, 317 (1972).

\bibitem{Ansorge:1988kn} 
R.~E.~Ansorge {\it et al.}  [UA5 Collaboration],
Z.\ Phys.\ C {\bf 43}, 357 (1989).

\bibitem{Aamodt:2010ft} 
K.~Aamodt {\it et al.}  [ALICE Collaboration],
Eur.\ Phys.\ J.\ C {\bf 68}, 89 (2010).

\bibitem{Khachatryan:2010nk} 
V.~Khachatryan {\it et al.}  [CMS Collaboration],
JHEP {\bf 1101}, 079 (2011).

\bibitem{ZajcPLB175}
W.~A.~Zajc,
Phys.\ Lett.\ B {\bf 175}, 219 (1986).

\bibitem{Dumitru:2012yr} 
A.~Dumitru and Y.~Nara,
Phys.\ Rev.\ C {\bf 85}, 034907 (2012).

\bibitem{Ugoccioni:2004wn} 
R.~Ugoccioni and A.~Giovannini,
J.\ Phys.\ Conf.\ Ser.\  {\bf 5}, 199 (2005)
[hep-ph/0410186];
D.~Prorok,
Int.\ J.\ Mod.\ Phys.\ A {\bf 26}, 3171 (2011)
[arXiv:1101.0787 [hep-ph]];
T.~Mizoguchi and M.~Biyajim,
arXiv:1207.0916 [hep-ph].

\bibitem{KNOhard}
A.~Bassetto, M.~Ciafaloni and G.~Marchesini,
Nucl.\ Phys.\ B {\bf 163}, 477 (1980);
Y.~L.~Dokshitzer, V.~S.~Fadin and V.~A.~Khoze,
Z.\ Phys.\ C {\bf 18}, 37 (1983);
Y.~L.~Dokshitzer,
Phys.\ Lett.\ B {\bf 305}, 295 (1993);
G.~P.~Salam,
Nucl.\ Phys.\ B {\bf 449}, 589 (1995).

\bibitem{KNV}
A.~Krasnitz, Y.~Nara and R.~Venugopalan,
Phys.\ Rev.\ Lett.\  {\bf 87}, 192302 (2001);
Nucl.\ Phys.\ A {\bf 727}, 427 (2003);
T.~Lappi,
Phys.\ Rev.\ C {\bf 67}, 054903 (2003).

\bibitem{ktF}
D.~Kharzeev, E.~Levin and M.~Nardi,
Nucl.\ Phys.\ A {\bf 730}, 448 (2004)
[Erratum-ibid.\ A {\bf 743}, 329 (2004)];
Nucl.\ Phys.\ A {\bf 747}, 609 (2005);
A.~Dumitru, D.~E.~Kharzeev, E.~M.~Levin and Y.~Nara,
Phys.\ Rev.\ C {\bf 85}, 044920 (2012);
J.~L.~Albacete, A.~Dumitru, H.~Fujii and Y.~Nara,
arXiv:1209.2001 [hep-ph].

\bibitem{Gelis:2009wh} 
F.~Gelis, T.~Lappi and L.~McLerran,
Nucl.\ Phys.\ A {\bf 828}, 149 (2009).

\bibitem{Tribedy:2010ab}
P.~Tribedy and R.~Venugopalan,
Nucl.\ Phys.\ A {\bf 850}, 136 (2011)
[Erratum-ibid.\ A {\bf 859}, 185 (2011)];
arXiv:1112.2445 [hep-ph].

\bibitem{Lappi:2009xa} 
T.~Lappi, S.~Srednyak and R.~Venugopalan,
JHEP {\bf 1001}, 066 (2010).

\bibitem{Schenke:2012hg} 
B.~Schenke, P.~Tribedy and R.~Venugopalan,
arXiv:1206.6805 [hep-ph].

\bibitem{quartic}
A.~Dumitru and E.~Petreska,
Nucl.\ Phys.\ A {\bf 879}, 59 (2012).

\bibitem{Iancu:2002aq} 
E.~Iancu, K.~Itakura and L.~McLerran,
Nucl.\ Phys.\ A {\bf 724}, 181 (2003);
A.~Dumitru, J.~Jalilian-Marian, T.~Lappi, B.~Schenke and R.~Venugopalan,
Phys.\ Lett.\ B {\bf 706}, 219 (2011);
E.~Iancu and D.~N.~Triantafyllopoulos,
JHEP {\bf 1111}, 105 (2011);
JHEP {\bf 1204}, 025 (2012).

\bibitem{rcBK}
I.~Balitsky,
Phys.\ Rev.\ D {\bf 75}, 014001 (2007);
Y.~V.~Kovchegov and H.~Weigert,
Nucl.\ Phys.\ A {\bf 784}, 188 (2007);
Nucl.\ Phys.\ A {\bf 789}, 260 (2007);
J.~L.~Albacete and Y.~V.~Kovchegov,
Phys.\ Rev.\ D {\bf 75}, 125021 (2007).

\bibitem{Iancu:2006jw} 
E.~Iancu, J.~T.~de Santana Amaral, G.~Soyez and D.~N.~Triantafyllopoulos,
Nucl.\ Phys.\ A {\bf 786}, 131 (2007).

\bibitem{Dumitru:2007ew}
A.~Dumitru, E.~Iancu, L.~Portugal, G.~Soyez and D.~N.~Triantafyllopoulos,
JHEP {\bf 0708}, 062 (2007).

\bibitem{Stasto:2000er} 
A.~M.~Stasto, K.~J.~Golec-Biernat and J.~Kwiecinski,
Phys.\ Rev.\ Lett.\  {\bf 86}, 596 (2001).

\bibitem{DuslingFernandezVenugopalan}
K.~Dusling, D.~Fernandez-Fraile and R.~Venugopalan,
Nucl.\ Phys.\ {\bf A828}, 161 (2009)
[arXiv:0902.4435 [nucl-th]].

\bibitem{rho_n_action}
S.~Jeon and R.~Venugopalan,
Phys.\ Rev.\ D {\bf 70}, 105012 (2004);
Phys.\ Rev.\ D {\bf 71}, 125003 (2005);
A.~Dumitru, J.~Jalilian-Marian and E.~Petreska,
Phys.\ Rev.\ D {\bf 84}, 014018 (2011)

\end{thebibliography}
\end{document}